\documentclass[useAMS,usenatbib]{mn2e}

\usepackage{graphicx}

\title[On the Origin of the Absorption Features in SS\,433]{On the Origin of 
the Absorption Features in SS\,433}
\author[A. D. Barnes et al.]{A. D. Barnes$^{1}$\thanks{E-mail:
adb@astro.soton.ac.uk (ADB)}, J. Casares$^{2}$, P. A. Charles$^{1,3}$, 
J. S. Clark$^{4,5}$, R. Cornelisse$^{1}$, 
\newauthor C. Knigge$^{1}$ and D. Steeghs$^{6}$\\
$^{1}$School of Physics and Astronomy, University of Southampton, SO17 1BJ, 
UK\\
$^{2}$Instituto de Astrofisica de Canarias, 38200 La Laguna, Tenerife, Spain\\
$^{3}$South African Astronomical Observatory, PO Box 9, Observatory 7935, 
Cape, South Africa\\
$^{4}$Department of Physics and Astronomy, University College London, Gower 
Street, London, WC1E 6BT, UK\\
$^{5}$Department of Physics and Astronomy, The Open University, Walton Hall, 
Milton Keynes, MK7 6AA, UK\\
$^{6}$Harvard-Smithsonian Center for Astrophysics, MS-67, 60 Garden Street, 
Cambridge, MA 02138, USA}
\begin{document}

\date{}

\pagerange{\pageref{firstpage}--\pageref{lastpage}} \pubyear{2002}

\maketitle

\label{firstpage}

\begin{abstract}
We present high-resolution optical spectroscopy of the X-ray binary system 
SS\,433, obtained over a wide range of orbital phases.  The spectra display 
numerous weak absorption features, and include the clearest example seen to 
date of those features, resembling a mid-A type supergiant spectrum, that have 
previously been associated with the mass donor star.  However, the 
new data preclude the hypothesis that these features originate solely within 
the photosphere of the putative mass donor, indicating that there may be more 
than one region within the system producing an A supergiant-like spectrum, 
probably an accretion disc wind.  Indeed, whilst we cannot confirm the 
possibility that the companion star is visible at certain phase combinations, 
it is possible that all supergiant-like features observed thus far are 
produced solely in a wind.  We conclude that great care must be taken when 
interpreting the behaviour of these weak features.
\end{abstract}


\section{Introduction}

\begin{figure*}
 \includegraphics[width=1.0\linewidth, clip]{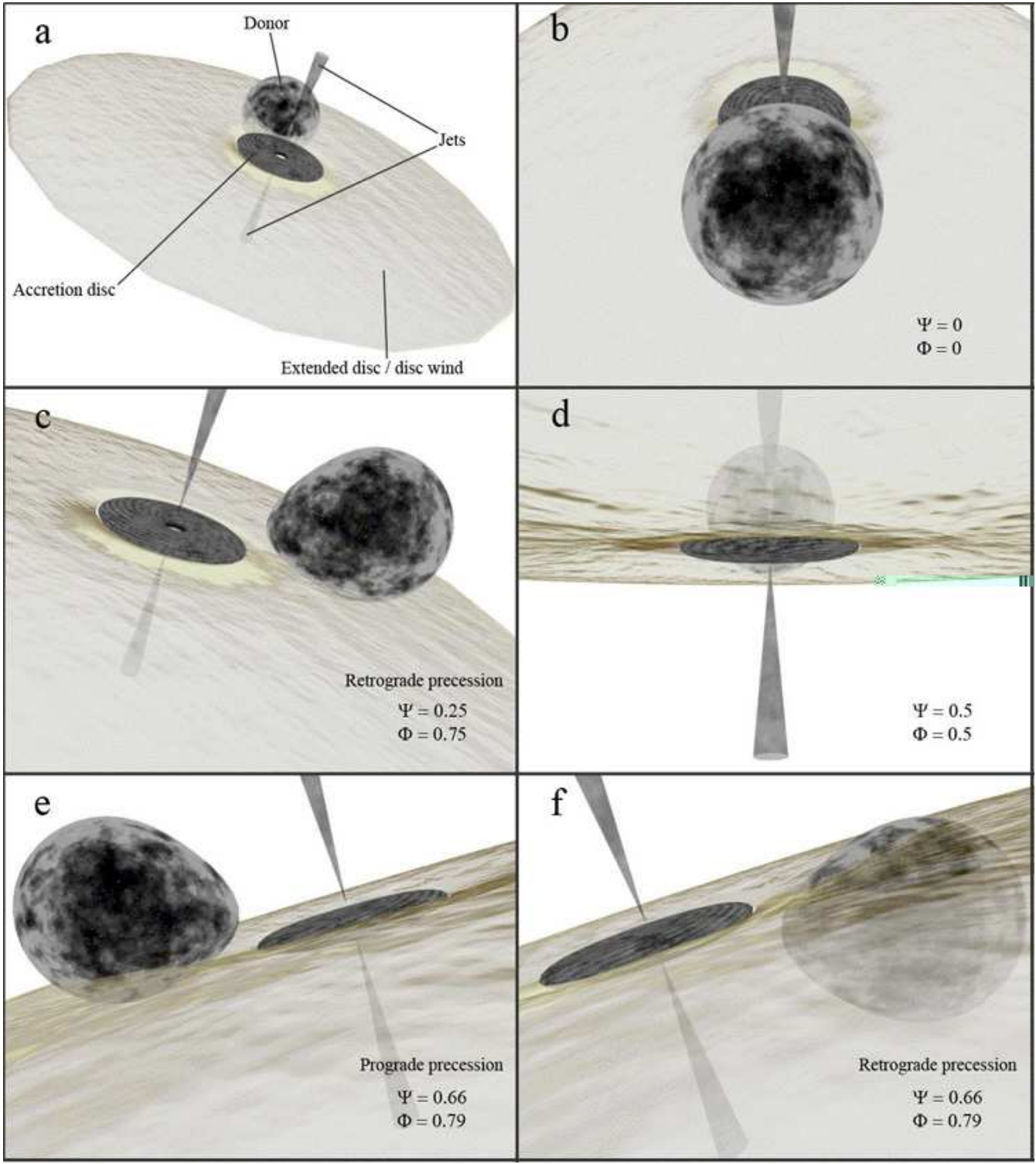}
 \caption{Cartoon picture of the SS\,433 system with an equatorially 
expanding extended disc, from the point of view of an observer on earth.  In 
all panels the sense of precession is represented as being anti-clockwise, 
with orbital motion clockwise in the retrograde case and anti-clockwise in the 
prograde case.  
{\it a}) Overview.  Note the extended disc is arbitrarily cut-off in the 
radial direction, and its opening angle is rather narrow, 
neither of which are confirmed physical parameters.  
{\it b}) Preferred phase combination of \protect\cite{hill}.  Donor could still be 
visible above the extended disc from $\phi_{orb}$ $\sim$ 0.75--0.25.  
{\it c}) Later $\Psi_{prec}$, but the donor could still be visible above the 
extended disc at this combination of phases.  
{\it d}) Donor star becomes fully masked behind the extended disc, but will 
become partially visible {\it below} the disc from $\phi_{orb} \sim$ 0.75--0.
25.  
{\it e}) \& {\it f}) The system alignment on 2004 June 29.  Here the donor is 
mostly visible assuming prograde precession, and masked in the case of 
retrograde (the preferred model).
}
 \label{cartoon}
\end{figure*}

The bizarre object SS\,433 is a galactic X-ray binary at the centre of the 
supernova remnant W\,50.  It is observed as a weak ($\sim$10$^{36}$erg~s$^
{-1}$) X-ray source (X1909+048), which combined with the high orbital 
inclination ({\it i} $\sim 78\degr$) and low L$_X$/L$_{opt}$ ratio ({\it V}
$\sim$14.2) suggests that it may be an accretion disc 
corona source.  In addition to its pair of relativistic 
jets (v $\sim$ 0.26c), recent radio observations have detected extended 
emission in a direction that is perpendicular to the main jet outflow 
\citep{bl}.  This provides evidence for some form of circumbinary outflow 
such as a disc-like outflow of matter from the outer parts of the accretion 
disc \citep{zw}.  The overall system geometry suggested 
by these recent studies is illustrated in Fig. \ref{cartoon}{\it a}, a 
schematic that we will use extensively throughout this paper.

The system displays two key periodicities, a $\sim$ 162d precession period of 
its jets and a $\sim 13$d orbital period.  The precession of the jets is 
revealed in the 
radial velocity curves of the strong, blue- and red-shifted 'moving' Balmer 
emission lines \citep{margon}.  This 162d modulation is also present in the 
optical flux, a result of the changing orientation of the precessing accretion 
disc.  The orbital period was first discovered through the radial 
velocity variations of the 'stationary' Balmer and He\,I emission lines 
\citep{cch}, which reach maximum velocity at superior conjunction of the 
X-ray source.  However, the He\,II $\lambda$4686 has a 
different radial velocity curve, with a velocity maximum occurring when the 
compact object is receding \citep{ch,fb}.  This may indicate an origin in or 
around the accretion disc/compact object.

Due to the strong and broad emission lines, the detection of any spectroscopic 
signatures of the donor star has proved rather more difficult, and the 
suggestions for its spectral type range from OB to Wolf-Rayet \citep{ch,
van}.  Since the interpretation of the radial velocity curves requires 
assumptions about the mass donor, estimates for the mass of the compact object 
have ranged as widely as 0.8 $M_{\sun}$ \citep{dod} and 62 $M_{\sun}$ 
\citep{ant}.  \cite{ghm} and \cite{hill} suggested that the optimal time 
to detect a signature of the donor star is during the X-ray eclipse 
($\phi_{orb}\sim0$), at a precessional phase when the disc is most open 
to the observer ($\Psi_{prec}\sim0$, Fig. \ref{cartoon}{\it b}).
  During this 
configuration of phases, light from the donor should be least 'obscured' 
by the extended circumbinary disc or disc wind.  Based on observations taken 
during this combination of phases, 
\cite{hill} discovered features corresponding with a mid-A supergiant (SG) 
and suggested that these could originate upon the donor star.  This 
constrained the binary system to be a low mass 
black hole (2.9 $\pm~ 0.7 M_{\sun}$) with a 10.9 $\pm~ 3.1 M_{\sun}$ 
companion, consistent with the predictions of \cite{king}.  

However, these calculations were based upon limited 
orbital phase coverage.  In this paper we will test the possibility of 
observing the A SG features over a much broader range of 
phases in order to confirm their origin and provide a more secure estimate 
of the system parameters.  In Section 2 we give an overview of the 
observations.  In Section 3 we will begin with a general description of all 
our spectra, then discuss the three best spectra in more detail (Sections 
3.1-3.3), and finish by presenting the radial velocities in Section 3.4.  We 
will begin Section 4 with a discussion of the consequences of the model 
proposed by \cite{ghm} and then discuss our findings.

\section{Observations}

SS\,433 was observed using three different telescopes and instruments
in the blue spectral range ($\lambda\lambda$4000-5000) between 2000
and 2004. 
Four high-resolution spectra of SS\,433 were obtained with the TWIN
spectrograph mounted on the 3.5~m telescope at Calar Alto observatory.
A 1200 lines mm$^{-1}$ grating together with a 1.5$\arcsec$ long slit covered
$\lambda\lambda$3700-4800 at 0.51 \AA\ pix$^{-1}$, yielding a spectral 
resolution of 67 km s$^{-1}$ (FWHM) at 4550\AA.  Two intermediate resolution 
spectra (0.63 \AA\ pix$^{-1}$)
were obtained using the Intermediate Dispersion Spectrograph (IDS)
attached to the 2.5~m Isaac Newton Telescope (INT) at the Observatorio del
Roque de Los Muchachos on the nights of 15-16 August 2002. We used the R900V 
grating in combination with the 235~mm
camera and a 1.2$\arcsec$ slit to provide a spectral resolution of 83 km
s$^{-1}$ (FWHM) at 4550\AA. 

Nine high resolution spectra of SS\,433 in addition to an A SG comparison 
star were also obtained using the blue arm of the ISIS spectrograph with the 
4.2~m William Herschel Telescope (WHT) at the Observatorio del
Roque de Los Muchachos in June/July 2004. The R1200B grating was used with a 
central wavelength setting of 4550 \AA\, and a 0.93$\arcsec$ slit, giving a 
spectral resolution of 29 km s$^{-1}$ (FWHM) 4550\AA.
A complete log of the observations is presented in Table \ref{tabobs}.  
Orbital phases have been calculated using the light-curve ephemeris of 
\cite{goran}, whilst disk precession phases are based on the ephemeris of 
\cite{G2k2}.

The images were bias corrected and flat-fielded, and the spectra
subsequently extracted using conventional optimal extraction techniques in
order to optimize the signal-to-noise ratio of the output \citep{Horne86}.
Frequent observations of comparison arc lamp images
were performed in the course of each run and the pixel-to-wavelength scale
was derived through polynomial fits to a large number of identified
reference lines. The final rms scatter of the fit was always $<$1/30 of
the spectral dispersion.  Finally, the spectra have been rectified using a 
high-pass filter to 
remove any low-order variations (0.001--0.08 cycles/\AA) caused by the broad 
jet lines, and then the data were combined to produce an average spectrum for 
each night.

\begin{table*}
\centering
\caption[]{Observing log of our SS\,433 spectroscopy, and derived parameters.  
For each observation we give the orbital phase, $\phi_{orb}$, and 
precessional phase, $\Psi_{prec}$, at the time of the observation.  We also 
give the radial velocities obtained via cross-correlation fitting, the 
estimated scaling factor of the absorption lines compared to a mid-A 
SG comparison star, {\it f}, and the optimal broadening applied to the 
comparison star, {\it v} sin {\it i} (see 
Sect. 3).  The spectra from 2000-07-23, 2004-07-02 and 2004-07-04 do not 
display clear A SG-like features.}
\label{tabobs}
\scriptsize
\begin{tabular}{lccccccccc}
\hline
\hline
Date    & HJD & Obs. & Exp. Time & Reciprocal Dispersion & $\phi_{orb}$ & $\Psi_{prec}$ &
Radial Velocity & {\it f} & {\it v} sin {\it i} \\
& -2450000 & & (seconds) &  (\AA\ pix$^{-1}$) & & & (km s$^{-1}$) & & 
(km s$^{-1}$) \\
\hline
2000-07-20 & 1746.38 & C.A. & 1800 & 0.51 & 0.687 & 0.777 & -54 $\pm$ 15 & 0.80 $\pm$ 0.09 & 97 $\pm$ 30 \\
2000-07-21 & 1747.43 & C.A. & 1800 & 0.51 & 0.768 & 0.784 & -91 $\pm$ 16 & 1.03 $\pm$ 0.10 & 110 $\pm$ 20 \\
2000-07-22 & 1748.36 & C.A. & 1800 & 0.51 & 0.839 & 0.789 & -20 $\pm$ 22 & 1.04 $\pm$ 0.05 & 108 $\pm$ 15\\
2000-07-23 & 1749.42 & C.A. & 1800 & 0.51 & 0.920 & 0.796 & - & - & - \\
2002-08-15 & 2501.64 & INT & 1800 & 0.63 & 0.420 & 0.435 & -120 $\pm$ 5 & 0.51 $\pm$ 0.05 & 80 $\pm$ 8 \\
2002-08-16 & 2502.55 & INT & 1800 & 0.63 & 0.490 & 0.441 & -115 $\pm$ 12 & 0.29 $\pm$ 0.07 & 77 $\pm$ 10 \\
2004-06-29 & 3186.67 & WHT & 3x1800 & 0.22 & 0.784 & 0.660 & 6 $\pm$ 2 & 0.77 $\pm$ 0.04 & 84 $\pm$ 5 \\
2004-06-30 & 3187.63 & WHT & 3x1800 & 0.22 & 0.857 & 0.666 & -5 $\pm$ 8 & 0.46 $\pm$ 0.08 & 98 $\pm$ 17 \\
2004-07-01 & 3188.67 & WHT & 3x1800 & 0.22 & 0.937 & 0.672 & 4 $\pm$ 13 & 0.65 $\pm$ 0.05 & 39 $\pm$ 4 \\
2004-07-02 & 3189.68 & WHT & 3x1600 & 0.22 & 0.014 & 0.678 & - & - & - \\
2004-07-03 & 3190.51 & WHT & 3x1800 & 0.22 & 0.077 & 0.683 & -12 $\pm$ 13 & 0.30 $\pm$ 0.06 & 53 $\pm$ 8 \\
2004-07-04 & 3191.63 & WHT & 3x1800 & 0.22 & 0.163 & 0.690 & - & - & - \\
2004-07-20 & 3207.43 & WHT & 6x600 & 0.22 & 0.371 & 0.788 & -78 $\pm$ 8 & 0.10 $\pm$ 0.06 & 73 $\pm$ 20 \\
2004-07-30 & 3217.43 & WHT & 3x1800 & 0.22 & 0.135 & 0.849 & -1 $\pm$ 5 & 0.59 $\pm$ 0.04 & 49 $\pm$ 4 \\
2004-07-31 & 3218.50 & WHT & 3x1800 & 0.22 & 0.217 & 0.856 & -35 $\pm$ 4 & 1.05 $\pm$ 0.08 & 117 $\pm$ 10$^*$ \\
\hline
\multicolumn{9}{l}{$^*$ Measurement of the combined features in the 
double-lined spectrum (see Sect. 3.3)}
\end{tabular}
\end{table*}

\section{Analysis}

We analysed a portion of the continuum that is free of stationary 
emission lines and rich in absorption features.  The best 
available region encompasses the range between $\sim$4500 -- 4630 \AA, 
which is bracketed by emission from He\,I $\lambda$4471 and the N\,III/C\,III 
complex $\sim$4640-4650 \AA.

\begin{figure*}
 \includegraphics[height=1.0\linewidth, clip,angle=-90]{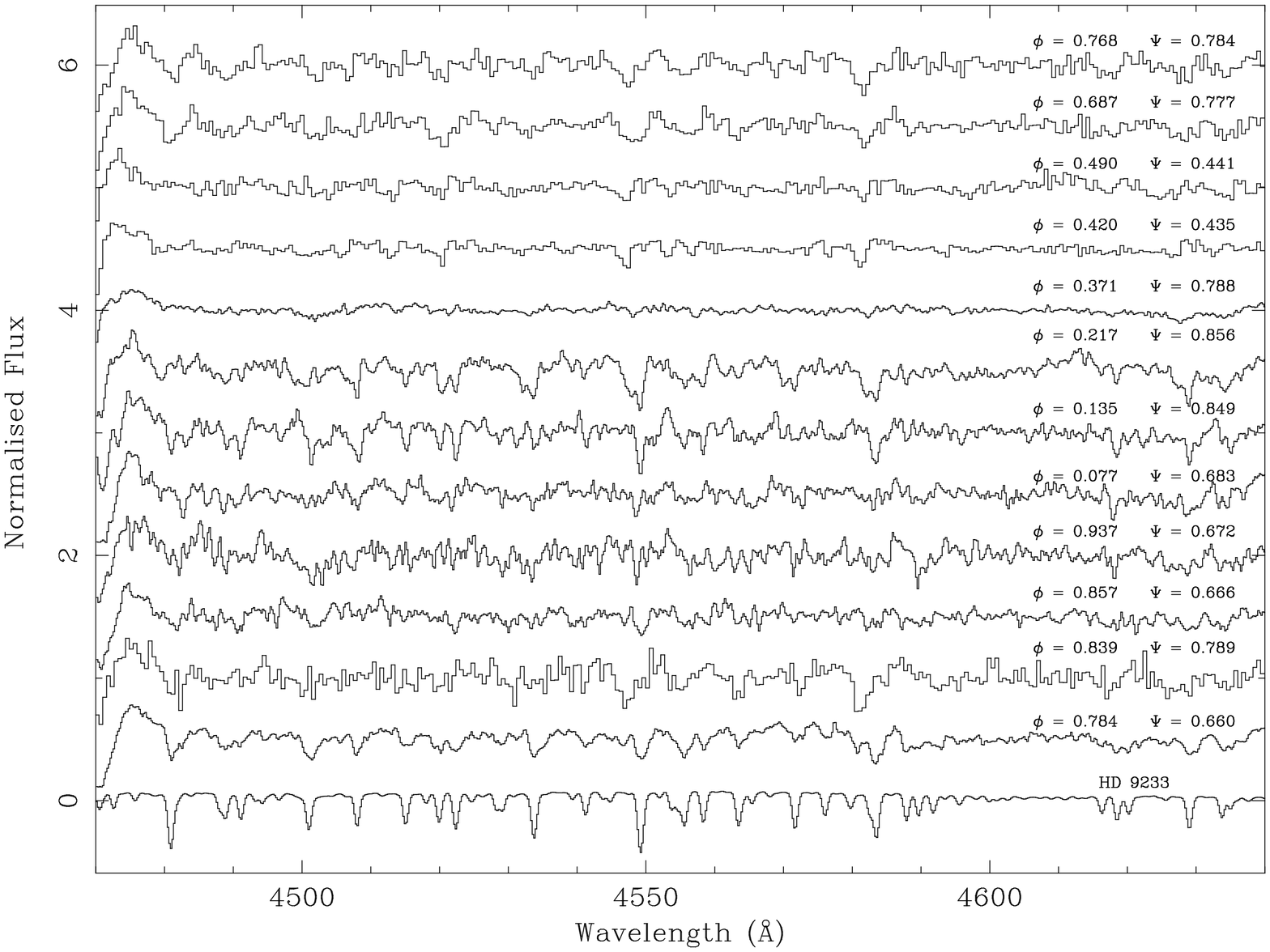}
 \caption{All of the nightly averaged spectra which display key A 
supergiant-like features arranged in orbital phase order beginning with the 
spectrum discussed in Section 3.1 (second from bottom).  The template star 
HD\,9233 is shown at the bottom.  Line 
identifications have been omitted here for clarity, see Figs. \ref{Jun29}, 
\ref{Jul30}, \ref{Jul31} for this information.}
 \label{allspec}
\end{figure*}

Spectra from twelve of the fifteen separate nights, covering a broad range of  
orbital phases, display photospheric features that appear to be similar to 
that of an A SG (Fig. 
\ref{allspec}).  Whilst the strongest Fe\,II absorption lines ($\lambda4549~ 
\&~ \lambda4584$ \AA) are present in each of these spectra, the morphology of 
these features, and indeed the presence of other weaker features, is highly 
variable, a fact which seems difficult to reconcile with the donor star being 
the only system component producing these lines.  

We have extracted velocity measurements from each of the spectra which 
display clear A SG-like photospheric features by cross-correlating with an 
optimally broadened comparison star (HD\,9233).  The systemic velocity of the 
comparison star was first removed ($\gamma$ = -34 $\pm$ 2 km s$^{-1}$, Hillwig 
et al. 2004) and then the data were re-binned into velocity space and 
cross-correlated with the optimally broadened spectrum of HD\,9233 across the 
range $\sim$ 4500 -- 4630 \AA. The comparison star was 
smoothed to the resolution of the spectra for the Calar Alto and INT runs 
using a Gaussian smoothing algorithm.  We present these results in Section 
3.4, after a more detailed description and comparison of three of these 
spectra which display the clearest absorption lines. 

For each nightly average spectrum, we broadened our A SG template from 39 to 
99 km s$^{-1}$ in steps of 2 km s$^{-1}$, taking into account that the 
template star HD\,9233 already possesses an intrinsic rotational broadening of 
39 km s$^{-1}$ \citep{hill}.  A spherical rotational profile \citep{gray} was 
applied, with a linear limb-darkening law of 
coefficient $\epsilon$ = 0.62, interpolated for $\lambda$ = 4500 \AA\ and 
{\it T}$_{eff}$ $\simeq$ 8350~K \citep{kbp,al}.  The broadened versions of the 
template star were multiplied by fractions {\it f} $<$ 1, to account for the 
fractional contribution to the total light, and subsequently subtracted from 
the nightly average spectra for SS\,433.  Minimising $\chi^2$ yields an 
optimal value for {\it f} and {\it v} sin {\it i} in each case (see Table 
\ref{tabobs}).

\subsection{A supergiant spectrum in SS\,433?}

\begin{figure}
 \includegraphics[height=1.0\linewidth, clip,angle=-90]{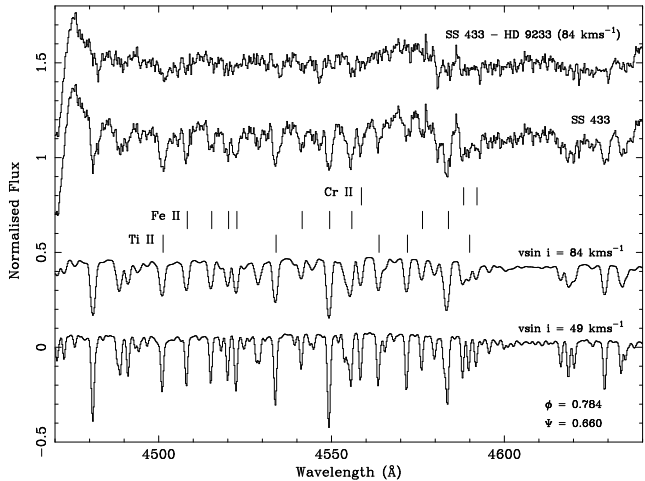}
 \caption{Average spectrum of SS\,433 from 2004 June 29 
($\Psi_{prec}\sim 0.66,~ \phi_{orb}\sim0.78$), 
together with the template star HD\,9233 broadened by different amounts 
(below), and the residual of SS\,433 - optimally broadened template (top).}
 \label{Jun29}
\end{figure}

As can be seen in Fig. \ref{Jun29} ($\Psi_{prec}\sim 0.66,~ 
\phi_{orb}\sim0.78$), we have a remarkable match between the SS\,433 spectrum 
and the A4\,Iab comparison star.  All of the key absorption lines present 
in this region of the comparison star also manifest themselves in our spectrum 
of SS\,433, though broader and somewhat shallower, as might be expected due to 
the continuum emission of the accretion disc.  A $\chi^2$ test on the 
residuals yields {\it v} sin {\it i} = 84 $\pm$ 5 km s$^{-1}$, whilst the 
depth of the lines are equivalent to an optimal scaling factor of 
0.77 $\pm$ 0.04.  Cross-correlating with the comparison star HD\,9233 results 
in a velocity shift measurement of 6 $\pm$ 2 km s$^{-1}$.

Subtracting the optimally broadened template star from the average spectrum 
of SS\,433 provides a remarkably clean residual spectrum (Fig. \ref{Jun29}, 
top), with the only unusual features in the region of interest 
being the He\,I $\lambda$4471 P Cyg profile, and a blue-shifted jet line bump 
(He\,I $\lambda$4713) which has not been fully eliminated during the high-pass 
filtering process.

\subsection{Narrow Absorption Lines}

\begin{figure}
 \includegraphics[height=1.0\linewidth, clip,angle=-90]{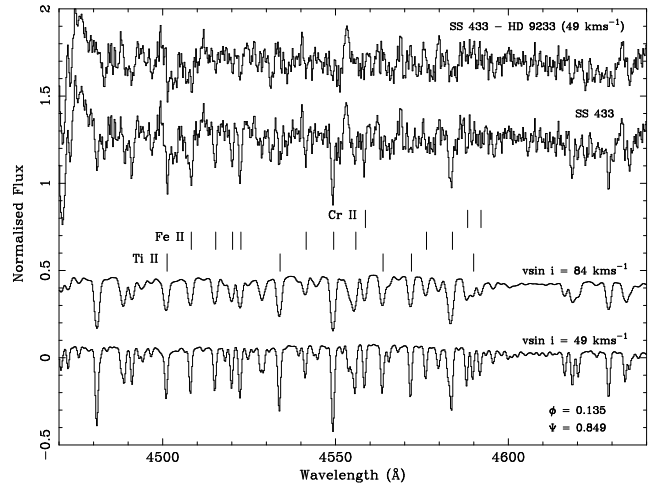}
 \caption{Average spectrum of SS\,433 from 2004 July 30 
($\Psi_{prec}\sim 0.85,~ \phi_{orb}\sim0.14$).  
The template star HD\,9233 is shown below for comparison, broadened by 
different amounts, and the residual of SS\,433 - optimally broadened template 
(top).  The emission feature at $\sim$ 4475\,\AA\ is a component of the He\,I 
$\lambda$4471 P Cyg profile (see text).}
 \label{Jul30}
\end{figure}

The spectrum displayed in Fig. \ref{Jul30} ($\Psi_{prec}\sim 0.85,~ 
\phi_{orb}\sim0.14$) is also very rich in the A SG-like 
features identified by \cite{hill}, including many Fe\,II, Cr\,II and Ti\,II 
lines.  However, it is immediately apparent even by eye 
that these absorption lines possess a much narrower and sharper profile (49 
$\pm$ 4 km s$^{-1}$ compared to 84 km s$^{-1}$) than those from the 
observation in Sect. 3.1.  

The fractional contribution of 0.59 $\pm$ 0.04 is somewhat lower too, 
indicating that a large proportion of the light in this spectral region, 
at this particular time, may be produced elsewhere in the system, e.g. the 
accretion disc.  

The family of absorption lines are slightly blue-shifted compared to the 
observation discussed in Sect. 3.1, with a velocity of -1 $\pm$ 5 km s$^{-1}$.
  If these lines are coming from the donor star then this 
is not the expected behaviour, which should have been close 
to its maximum blue-shift in the earlier observation, and approaching maximum 
red-shift at this orbital phase.

\subsection{Double Absorption Lines}

\begin{figure}
 \includegraphics[height=1.0\linewidth, clip,angle=-90]{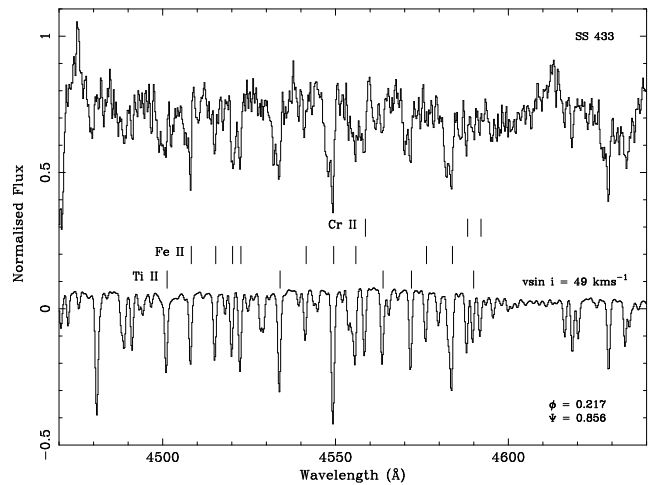}
 \caption{Average spectrum of SS\,433 from 2004 July 31 
($\Psi_{prec}\sim 0.86,~ \phi_{orb}\sim0.22$).  Note in particular the 
doubled absorption lines of Fe\,II and Ti\,II.  Again, the template star 
HD\,9233 is shown for comparison below, broadened to the optimal value from 
the previous night's observations.}
 \label{Jul31}
\end{figure}

Even more anomalously, the spectrum from the following night ($\Psi_{prec}\sim 
0.86,~ \phi_{orb}\sim0.22$, Fig. \ref{Jul31}) appears 
to possess absorption lines which are doubled (Fe\,II and Ti\,II), with a 
narrower red component blended with a 
broader, shallower blue.  This feature is evident in each of the individual 
spectra from the night, and is not therefore the signature of some systematic 
error introduced when producing the average spectrum for the night.  

Cross-correlating the data from this night against the A SG comparison star 
gives a velocity of -35 $\pm$ 4 km s$^{-1}$.  This is again much bluer than 
expected, with the observation coinciding with the expected maximal red-shift 
of the donor star.  
However, the cross-correlation technique in this case fails to fit either of 
the apparent absorption components in this spectrum, instead picking out an 
'in-between' velocity.  Gaussians fitted individually to the red and 
blue features indicate velocities of -10 $\pm$ 10 and -100 $\pm$ 15 km 
s$^{-1}$.  Neither component therefore possesses a velocity commensurate with 
the predicted behaviour of the donor star.

\subsection{Absorption Line Velocities}

\begin{figure}
 \includegraphics[width=1.0\linewidth, clip]{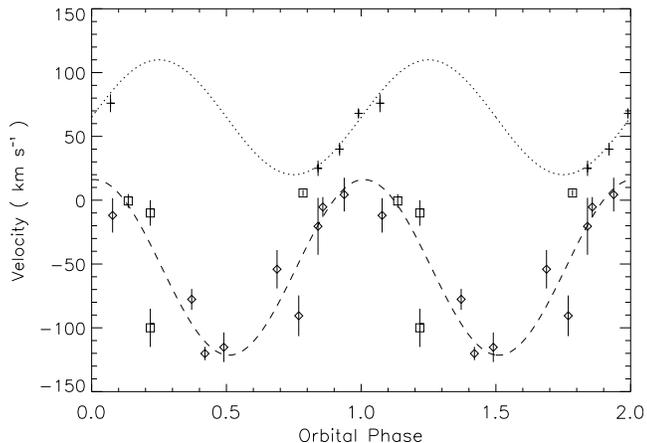}
 \caption{The radial velocities obtained via cross-correlation fitting in the 
$\lambda$4500--4600 \AA\ region.  The results obtained by \protect\cite{hill} are 
shown marked by pluses, together with their best-fitting sine curve (dotted 
line).  The remaining points are the results of our own campaign, with those 
marked by squares representing the spectra selected for discussion 
in Section 3.  The dashed line represents the best-fitting sine curve 
to these data.  Note that two orbits have been plotted for clarity.}
 \label{vels}
\end{figure}

Our data cover a broad range of orbital phases, and are displayed in Fig. 
\ref{vels} as diamonds.  The three spectra discussed in Sections 3.1--3.3 
are marked with squares.  Note that all the velocities used in this analysis 
were the result of cross-correlation with a comparison star, with the 
exception of the individual line fits discussed in Section 3.3.  The 
results reported by \cite{hill} are indicated by pluses, with the dotted line 
representing their best-fitting sine curve to these data.  The best-fit sine 
curve 
to our data is marked by a dashed line, with semi-amplitude K = 69 $\pm$ 4 
km s$^{-1}$, systemic velocity $\gamma$ = -53 $\pm$ 3 km s$^{-1}$ and with 
velocity maxima occuring at orbital conjunction, 0.26 $\pm$ 0.01 out of 
phase with the motion of the accretion disc \citep{fb}.

\section{Discussion}

\subsection{A model for SS\,433}

\cite{ghm} and \cite{hill} state that the mass donor is ideally observed 
around the primary eclipse and when the accretion disc has precessed to its 
most face-on position ($\Psi_{prec}\sim\phi_{orb}\sim0$). The donor star would 
then have the highest chance of being positioned above and between any 
extended disc and an observer (Fig. \ref{cartoon}{\it b}).  Extending this 
idea, it should also be possible to detect the donor star at certain other 
combinations of phases.  In Fig. \ref{cartoon}{\it c} we show an example, with 
the companion being positioned clearly above the extended disc, whilst in Fig. 
\ref{cartoon}{\it d} we 
have the inverse situation, with the companion star being totally obscured 
behind the extended disc.  However, at this $\Psi_{prec}$ the companion could 
become partially visible {\it below} the extended disc from $\phi_{orb} 
\sim$ 0.75--
0.25.  The extent to which we may be able to observe the companion will be 
determined by the opening angle and radial size of the extended disc, 
parameters which are currently not well constrained, in addition to the degree 
of irradiation of the inner face of the donor star from the accretion disc 
\citep{cher}.

If this scenario is correct, it can provide direct observational 
information about the precession evident in the system.  Confirming whether 
it be prograde or the preferred retrograde model \citep{news} with respect 
to the donor star could perhaps cast further light upon the 
physics driving the precession \citep{katz,leib}.  From Figs. \ref{cartoon}
{\it e} \& {\it f}, representing the system alignment of the spectrum 
discussed in Section 3.1, it 
becomes clear that the companion star could only be visible above the disc at 
this combination of phases if the precession were prograde in nature, with the 
extended disc masking the donor in the case of retrograde precession.

The presence of the same A SG-like features reported by \cite{hill} in many 
of our spectra, at a range of different orbital and precessional phases, would 
at first glance seem to indicate that our data support this scheme.  In 
particular, the 
observation discussed in Section 3.1 (where the accretion disc is close to 
edge-on, $\Psi_{prec}\sim 0.66,~ \phi_{orb}\sim0.78$) provides an unsurpassed 
match between an A4\,Iab comparison star and SS\,433 (Fig. \ref{Jun29}).  If 
these features really do originate upon the donor star, then the depth and 
strength of the lines are equivalent to a scaling factor of 0.77 $\pm$ 
0.04.  This would indicate a surprisingly high contribution to the total light 
at this combination of phases, compared to observations with the disc face-on, 
where \cite{hill} find a fractional contribution of 0.36 $\pm$ 0.07.  

We can compare our estimated scaling factor with that expected from 
observed light curves of SS\,433 in a similar manner to \cite{hill}.  The key 
assumption required is that the donor star is totally eclipsed by the disc at 
$\Psi_{prec} = 0$ and $\phi_{orb} = 0.5$, and at all other times both the 
donor and the accretion disc are contributing to the total light.  The 
{\it V}-band light-curves of 
\cite{zw} imply a potential fractional contribution for the donor star of 
roughly 0.70 $\pm$ 0.15 at $\Psi_{prec}\sim 0.66,~ \phi_{orb}\sim0.78$.  
Whilst this prediction appears to be consistent with our observational data, 
we must note that there is a large degree of scatter evident in the 
observed precessional light curves, introducing a degree of systematic error 
into these calculations.  Indeed, if we use the same method to predict the 
donor contribution at phase combination $\Psi_{prec}\sim\phi_{orb}\sim0$ from 
the \cite{zw} light curves, the resulting estimate is too high by a factor of 
three.


If we presume the absorption features noted in Section 3.1 to be a signature 
of 
the donor star, then retrograde precession may be immediately ruled out.  
However, the same features also occur in the spectra discussed in Sections 
3.2--3.3.  These observations were performed at a phase combination where the 
donor should be obscured assuming prograde precession, and visible in the 
retrograde case.  In addition to this apparent conflict, we must also account 
for the unusual velocity behaviour of these weak absorption features.

\subsection{A disc wind origin?}

Cross-correlating our spectra of SS\,433 with an A supergiant 
spectrum produces a velocity curve (Fig. \ref{vels}, $\gamma$ = -53 $\pm$ 3 km 
s$^{-1}$) which is far removed in terms of systemic velocity from the value of 
65 $\pm$ 3 km s$^{-1}$ reported by \cite{hill}, though it is close to the 
result of \cite{ghm}, where $\gamma$ =  -44 $\pm$ 9 km s$^{-1}$.  
Additionally, it is almost exactly a quarter out of phase with the \cite{fb} 
He\,II curve (where $\gamma$ = -22 $\pm$ 14 km s$^{-1}$) which traces the 
motion of the compact object.  
  

The occurence of a velocity maximum at orbital conjunction essentially 
precludes the possibility that these features were produced in the 
photosphere of the companion.  Whilst it is extremely difficult to pin down 
a particular region within the system which could produce this velocity 
signature, it is interesting to note that this behaviour has been reported 
before.  \cite{G2k2} find the 'stationary' He\,I and H$\alpha$ 
emission lines display a remarkably similar semi-amplitude and phasing to 
the A SG-like features investigated here.  This is 
attributed to a clumpy wind outflow from the accretion disc rather than the 
disc itself or to a gas stream \citep{gor,G2k2}.  Although there is a large 
disparity in line widths between the stationary emission lines and our narrow 
absorption lines, if what we observe in SS\,433 is a structured outflow then
we would expect to see narrow absorption components from localised density
enhancements in the wind.  This would appear 
to indicate that for at least some combinations of orbital and precessional 
phases the A SG 'photospheric' features arise not from the donor but instead 
from a localised area that can be associated with the region where the 
He\,I and H$\alpha$ stationary emission lines are produced.  

We also have some longer wavelength data collected during the 2004 observing 
campaign in La Palma with ISIS ($\sim$ 8400-8900\AA), to be published in a 
later paper.   Here we 
identify heavily blended Ca\,II and weak N\,I absorption lines, which again 
correspond to the absorption features expected for an early-mid A SG.  
Interestingly, there appears to be a strong positive correlation
between the strength of the P Cygni profiles seen in the Paschen lines and
the strength of the Ca\,II and N\,I lines, implying another link 
between traditional wind features and those typical of the proposed companion 
star.

The strongest evidence to date for an A SG donor is the data presented 
by \cite{hill}.  They base their identification on three criteria which 
suggest an association 
with the donor, namely the radial velocity curve characteristics, the 
absorption line depths and the absorption line width.  However, their data do 
not cover the key phases when there should be a turnover in velocity space 
($\phi_{orb}\sim$ 0.25, 0.75), which would unambiguously associate the lines 
with the donor.  Interestingly, a number of our own spectra which can be 
clearly associated with an outflow also meet at least two of the above 
criteria (see Table \ref{tabobs}), with only the velocity curve excluding an 
association with the donor. 


The large offset between the systemic velocity detected by \cite{hill} and in 
our data could be a potential problem for a purely circumbinary outflow 
origin.  However, this velocity offset could be explained by a 
considerable cycle to cycle variation in the semi-amplitude 
of the 'stationary' emission lines originating in the accretion disc wind 
\citep{G2k2}.
  Another problem could be the doubled absorption lines shown in Fig. 
\ref{Jul31}.  This could perhaps be 
indicative of two entirely separate regions with identical spectral signatures,
 though neither component corresponds to a possible donor star origin.  It 
seems more likely that two different components of a clumpy wind outflow could 
concurrently pass through the line-of-sight, as proposed by \cite{G2k2} and 
\cite{lep}.

\section{Conclusions}

Our series of high-resolution optical spectra of SS\,433 clearly display 
numerous typical 
A SG photospheric features, as reported by several authors and attributed to 
a putative mass donor.  However, the velocities of these features do
not correspond to the expected motion of the mass donor, but rather
to the H$\alpha$ and He\,I 'stationary' emission lines noted by \cite{G2k2} 
and assumed to originate in a clumpy, accretion-driven outflow.

Although the observations of \cite{hill} are suggestive of an A 
SG-like donor, we cannot exclude the possibility that these features originate 
from the same highly variable accretion-driven wind.  If these 
lines are consistent over a number of observations at $\Psi_{prec}\sim
\phi_{orb}\sim0$ then it would seem likely that they are a signature of the 
donor.  In this case the large off-set in systemic velocity between the 
\cite{hill} data and ours could be due to the velocity of the outflow towards 
our line of sight.  However, if the lines originate 
exclusively in some form of outflow, then we would expect them to be rather 
variable in terms of width, depth, shape and velocity.  In any case, 
considerable care must be taken to correctly interpret these features and 
thus determine the nature of the stellar components within the SS\,433 system.

\section*{Acknowledgments}

We gratefully acknowledge A.B. Hill and D. Moss for artistic direction, and Ph.
 Podsiadlowski and the referee D.R. Gies for helpful comments.  ADB is 
supported by a 
PPARC Studentship.  JSC is supported by an RCUK Fellowship.  DS acknowledges 
a Smithsonian Astrophysical Observatory Clay Fellowship.  JC acknowledges 
support from the Spanish MCYT grant AYA2002-0036.  Based on 
observations collected at the Centro Astronomico 
Hispano Aleman (CAHA) at Calar Alto, operated jointly by the 
Max-Planck Institut fur Astronomie and the Instituto de Astrofisica 
de Andalucia (CSIC).  The INT and WHT are operated on the island of La Palma 
by the Isaac Newton Group in the Spanish Observatorio del Roque de los 
Muchachos of the Instituto de Astrofisica de Canarias.  We gratefully 
acknowledge the use of the {\sc molly} package developed by T.R. Marsh.

\label{lastpage}

\end{document}